\documentclass[aps,pra,floatfix,reprint, superscriptaddress]{revtex4-1}

\usepackage{graphicx}
\usepackage{color}
\usepackage[normalem]{ulem}
\usepackage{hyperref} 
\usepackage{amsmath}

\begin{document}

\title{Time-variant metasurfaces enable tunable spectral bands of negative extinction}

\author{Maxim~R.~Shcherbakov}
\affiliation{School of Applied and Engineering Physics, Cornell University, Ithaca, NY 14853, USA}

\author{Robert~Lemasters}
\affiliation{Department of Physics, Emory University, Atlanta, GA 30322, USA}

\author{Zhiyuan~Fan}
\affiliation{School of Applied and Engineering Physics, Cornell University, Ithaca, NY 14853, USA}

\author{Jia~Song}
\affiliation{Department of Chemistry, Emory University, Atlanta, GA 30322, USA}

\author{Tianquan~Lian}
\affiliation{Department of Chemistry, Emory University, Atlanta, GA 30322, USA}

\author{Hayk~Harutyunyan}
\affiliation{Department of Physics, Emory University, Atlanta, GA 30322, USA}

\author{Gennady~Shvets}
\affiliation{School of Applied and Engineering Physics, Cornell University, Ithaca, NY 14853, USA}

\date{\today}

\begin{abstract}
We demonstrate that rapidly switched high-Q metasurfaces enable spectral regions of negative optical extinction. 
\end{abstract}

\maketitle

The extinction coefficient of light passing through any photonic structure is defined as the deficit of optical power in transmission. Because optical extinction is caused by absorption, reflectance, and scattering, it must be positive as long as the structure is composed of gain-free time-invariant material. Moreover, for dispersive materials and photonic structures, the frequency-dependent extinction coefficient $E(\lambda) \equiv 1-T(\lambda)>0$ must be positive for every wavelength $\lambda$ of the incident light, where $T(\lambda)$ is the transmission coefficient. 

This situation can be dramatically altered by time-variant media whose optical properties (e.g., their refractive index) is an explicit function of time. To illustrate this effect, we calculate the extinction of an incident optical pulse with complex-valued amplitude $s^+(t)$ by a single-mode time-varying metasurface (TVM) characterized by its amplitude $a(t)$,  natural frequency〖$\omega_0$, time-dependent damping rate $\gamma(t)$, and the radiative coupling rate $\gamma_r$.  Within the framework of coupled-mode theory (CMT) \cite{A}, the transmitted wave $s^-(t)$ is calculated according to the equations below:
\begin{align}
&\dot a(t)+i\omega_0 a(t)+[\gamma_{\rm r}+\gamma_{\rm nr} (t)]a(t)=\sqrt{\gamma_{\rm r}}  s^+(t), \\
&s^-(t)=s^+ (t)-\sqrt{\gamma_{\rm r}}  a(t),
\end{align}
where $\gamma_{\rm nr}(t)=\gamma(t)-\gamma_{\rm r}>0$ is a non-radiative damping rate of the mode. When the quality factor $Q(t)\equiv\omega_0/2[\gamma_{\rm r}+\gamma_{\rm nr} (t)]$ rapidly decreases from its high initial value of $Q_{\rm i}\gg1$ to its final value of $Q_{\rm f}<Q_{\rm i}$ due to rapid increase of non-radiative losses, the TVM is assumed to be Q-switched.
 
For a Gaussian input signal $s^+ (t)=s_0  \exp{(-i\omega_0 t-t^2/\tau_{\rm probe}^2)}$ incident on an instantaneously Q-switched TVM (from $Q_{\rm i}=100$ to $Q_{\rm f}=5$ at $t=\tau$), the mode evolution $a(t)$ is plotted in Fig. 1(a). We assume the following physical pulse and TVM parameters: $\omega_0=2\pi c/\lambda_0$ (where $\lambda_0=3.3$~$\mu$m and $c=3\times10^{10}$~cm/s is the speed of light), and $\tau=200$~fs. The transmission spectrum $T(\lambda,\tau)\equiv|s^- (\lambda,\tau)|^2/|s^+ (\lambda)|^2$, which depends on the Q-switching time $\tau$, and extinction $E(\lambda,\tau)\equiv 1-T(\lambda,\tau)$ are calculated. The latter is plotted in Fig.1(b). While $E(\lambda,\tau)>0$ for a metasurface that does not vary during the trapping time of the pulse (dashed line), negative extinction (NE) spectral regions corresponding to  $E(\lambda,\tau)<0$ emerge for a TVM with $\tau<Q_0/\omega_0$. The spectral spacing $\Delta\lambda\approx\lambda_0^2/2\pi c \tau$ between the NE regions is determined by the delay $\tau$ between TVM excitation at $t=0$ and switching at $t=\tau$.

\begin{figure}[b]\label{fig1}
\includegraphics[width=\columnwidth]{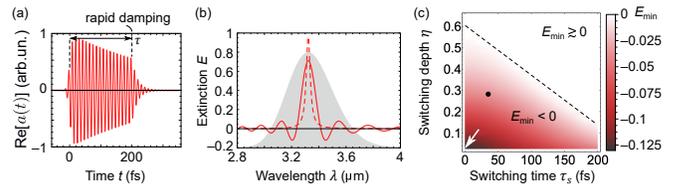}
\caption{Coupled-mode theory of negative light extinction by a TVM. (a) Evolution of the mode amplitude before and after the abrupt Q-switching at $t=\tau$ from $Q_{\rm i}=100$ to $Q_{\rm f}=5$ ($\eta=0.05$).  (b) Extinction spectra $E(\lambda)$ for $\tau=200$~fs (solid line) and $\tau=\infty$ (dashed line: no switching). Redistribution of spectral components manifests in several regions of $E(\lambda,\tau)<0$. Shaded area: spectrum of the incident pulse $s^+ (\lambda)$. (c) Conditions for negative extinction $E(\lambda,\tau)<0$:  Q-switching must be fast (horizontal axis) and deep (vertical axis). Switching time has been optimized to reduce the minimum extinction. The white arrow indicates the parameters used for panels (a,b), and the black dot shows the experimental parameters.
\label{fig1}}
\end{figure}

In a more realistic CMT calculation, we have used finite switching times $\tau_{\rm s}<\tau$ to establish the conditions for the emergence of the NE spectral region. According to Fig.1(c), $E(\lambda,\tau)<0$ is achieved for at least one value of $\tau$, as long as the switching is fast ($\tau_{\rm s}\ll \gamma_{\rm r}^{-1}$) and deep (small $\eta\equiv Q_{\rm f}/Q_{\rm i}$).

To experimentally demonstrate the NE phenomenon, we have chosen a TVM comprised of an array of subwavelength resonators (see Fig.2(a) for a scanning electron microscope image) made of amorphous germanium (a-Ge). Other high-Q photonic structures, such as microring and photonic crystal lattice defect resonators \cite{AA,AAA,AAAA,AAAAA}, as well as fiber-based ring cavities \cite{AAAAAA}, have been used to redistribute spectral components of light at the nanosecond and picosecond time scales. Recently, time-variant semiconductor metasurfaces have been used to generate new light frequencies over a broad spectral range \cite{AAAAAAA,AAAAAAAA}, revealing their promise for ultrafast operation and spatiotemporal shaping \cite{AAAAAAAAA}. To our knowledge, the NE phenomenon has not been reported. Here, we employ a TVM to demonstrate negative optical extinction in the mid-infrared (MIR) spectral range on a femtosecond time scale.

\begin{figure}\label{fig2}	
\includegraphics[width=1\columnwidth]{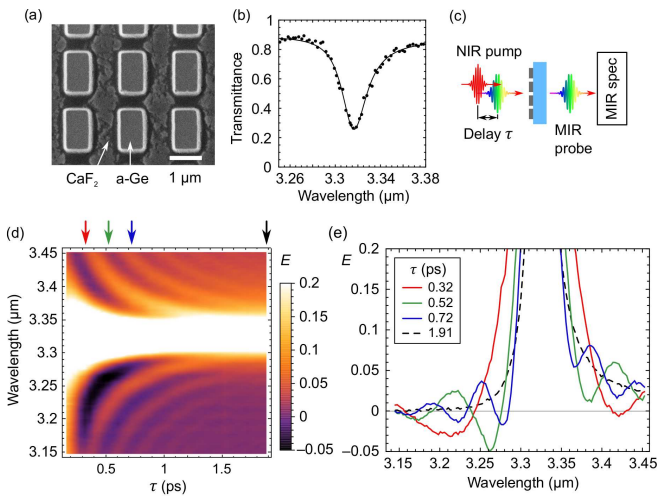}
\caption{Experimental realization of  negative optical extinction using semiconductor TVMs. (a) Scanning electron microscopy image of the (a-Ge)-based TVM. The a-Ge layer thickness is 200~nm. (b) Transmittance of the unperturbed resonant TVM: experimental data (dots) and a Fano-resonance fit (solid line). (c) Experimental schematic: a TVM interacting with a MIR probe is Q-switched by a NIR pump via electron-hole generation. Extinction spectra of the MIR radiation are detected by a spectrometer as a function of delay between the NIR and MIR pulses. (d) Experimental measurements of $E(\lambda,\tau)$ for different pump--probe delay times $\tau$. (e) Selected extinction spectra for $\tau=0.32$~ps (red), $\tau=0.52$~ps (green), $\tau=0.72$~ps (blue), and $\tau=1.92$~ps (black dashed).  Pump and probe parameters: durations $\tau_{\rm probe}=\tau_{\rm pump}=35$~fs, wavelengths $\lambda_{\rm probe}\approx \lambda_0$ and $\lambda_{\rm pump}=0.78$~$\mu$m, fluences $F_{\rm probe}=1$~$\mu$J/cm$^2$ and $F_{\rm pump}=150$~$\mu$J/cm$^2$.}\label{fig2}
\end{figure}

The TVM was designed to exhibit a sharp transmittance dip with a quality factor of $Q_{\rm i}\approx100$ at the resonant MIR wavelength $\lambda_0$ as shown in Fig.~2(b). A spectroscopic pump--probe apparatus sketched in Fig.~2(c) was used to study the optical extinction of a short MIR pulse by a TVR that was Q-switched by a delayed near-infrared (NIR) pulse of the same duration (see Fig.~2 caption for parameters). Single-photon pump absorption Q-switches the TVM over the $\tau_{\rm s}=\tau_{\rm pump}$ time via electron-hole plasma generation. Next, the differential transmission spectra $\Delta T (\lambda,\tau)/T(\lambda)$ (where $\Delta T(\lambda,\tau)\equiv T(\lambda,\tau)-T(\lambda,\infty)$ were measured in the pump--probe configuration) were converted into the extinction spectra $E(\lambda,\tau)=1-[T(\lambda)+\Delta T(\lambda,\tau)]/T_0$ using a Fano-resonance fit \cite{AAAAAAAAAA} of $T(\lambda)$ (Fig.~2(b), solid line). Note that the coefficient $T_0=0.89$ was applied to adjust the data for the residual reflectance that is not related to the properties of the metasurface. In contrast with the recently discovered blue-shifting of the entire spectrum due to the plasma driven increase of the refractive index \cite{AAAAAAAA}, here, the spectral reshaping happens on both the red and blue sides of the TVM resonance as shown in Fig. 2(d), where extinction spectrum is color-coded as a function of the delay time. In Fig. 2(e), those are replotted for four delay times. Indeed, as predicted in Fig. 1, at least one MIR spectral range corresponding to $E(\lambda,\tau)<0$ can be identified for all but the largest time delays; the latter corresponds to the effectively stationary metasurface. We have found good qualitative agreement between the experiment and theory for both cases of static metasurface (Fig. 1(b) dashed line and Fig. 2(e) dashed line) and the TVM (Fig. 1(b) solid line and Fig. 2(e) red line), suggesting NE regime is indeed realized in our TVM. Appreciable NE is found throughout the 3.14~$\mu$m~$<\lambda<3.28$~$\mu$m spectral range and is likely to extend further toward shorter wavelengths outside the detection range of our setup. Crucially, the spectral locations of the NE are tunable, controlled by the time delay $\tau$. The minimal measured extinction coefficient $E_{\rm min}^{\rm (exp)}\approx-0.05$ (Fig.2(e): green curve) is in good agreement with the theoretical value of $E_{\rm min}^{\rm (th)} \approx -0.03$  (Fig.1(c): black dot)  corresponding to the estimated experimental parameters of  $\eta^{\rm (exp)}\approx0.28$ and $\tau_{\rm s}=35$~fs.
 
In conclusion, we have predicted and experimentally demonstrated the existence of negative optical extinction of a femtosecond mid-infrared laser pulse in a time-variant semiconductor metasurface. Negative extinction by ultrafast control of the non-radiative loss in ultrathin resonators opens new opportunities for spectral shaping of short light pulses, in particular for amplifying its specific spectral components. The demonstrated approach can be potentially extended to other spectral ranges.

\end{document}